# Cyberbullying Detection: Exploring Datasets, Technologies, and Approaches on Social Media Platforms

ADAMU GASTON PHILIPO, School of Computer and Communication Engineering, University of Science and Technology Beijing, China
DOREEN SEBASTIAN SARWATT, School of Computer and Communication Engineering, University of Science and Technology Beijing, China
JIANGUO DING, Department of Computer Science, Blekinge Institute of Technology, 371 79, Sweden
MAHMOUD DANESHMAND, Department of Business Intelligence and Analytics and the Department of Computer Science, Stevens Institute of Technology, USA
HUANSHENG NING, School of Computer and Communication Engineering, University of Science and Technology Beijing, China

Cyberbullying has become a major challenge in the digital era, and many people, especially adolescents, use social media platforms to communicate and share information. Some exploit these platforms to embarrass others through messages, emails, speech, and public posts, causing severe psychological harm to victims. This study reviews existing research on technologies, approaches, datasets, and evaluation metrics for cyberbullying detection, while highlighting future directions and key challenges. The findings show that traditional models work reasonably well with small datasets but require constant updates; machine learning models face feature extraction and linguistic limitations; deep learning models perform better but lack multilingual and cross-lingual capabilities; and large language models (LLMs) achieve the highest performance, offering flexibility and rich linguistic features but face issues of high-energy use and real-time applicability. Addressing technological, methodological, dataset, and linguistic challenges will improve cyberbullying detection, helping to protect online communication and promote social responsibility.

---

Authors' Contact Information: Adamu Gaston Philipo, School of Computer and Communication Engineering, University of Science and Technology Beijing, Beijing, China; Doreen Sebastian Sarwatt, School of Computer and Communication Engineering, University of Science and Technology Beijing, Beijing, China; Jianguo Ding, Department of Computer Science, Blekinge Institute of Technology, 371 79, Karlskrona, Sweden; Mahmoud Daneshmand, Department of Business Intelligence and Analytics and the Department of Computer Science, Stevens Institute of Technology, Hoboken, NJ, USA; Huansheng Ning, School of Computer and Communication Engineering, University of Science and Technology Beijing, Beijing, China.

---

| Acronym | Description |
|---|---|
| API | Application Programming Interface |
| LLMs | Large Language Models |
| ANN | Artificial Neural Network |
| SVM | Support Vector Machine |
| KNN | K-Nearest Neighbors |
| MLP | Multilayer Perceptron |
| DBN | Deep Belief Network |
| DNN | Deep Neural Network |
| CNN | Convolutional Neural Network |
| RNN | Recurrent Neural Network |
| FNN | Feedforward Neural Network |
| GPT | Generative Pre-trained Transformer |
| BERT | Bidirectional Encoder Representations from Transformers |
| GRU | Gated Recurrent Unit |
| BiGRU | Bidirectional Gated Recurrent Unit |
| LSTM | Long Short-Term Memory |
| BiLSTM | Bidirectional Long Short-Term Memory |
| BiRNN | Bidirectional Recurrent Neural Network |
| GAN | Generative Adversarial Network |
| RBM | Restricted Boltzmann Machine |
| TF-IDF | Term Frequency-Inverse Document Frequency |
| BoW | Bag of Words |
| PoS | Part of Speech |

## 1 INTRODUCTION

The participation of people in social media platforms is increasing rapidly because it is the simplest way to communicate and share information [1]. There are a large number of users on social media platforms such as Facebook, YouTube, Instagram, and Twitter [2]. Social media platforms allow users to share information by sending text, photos, videos, audio, and animations about cyberbullying [3]. The total world population is estimated at 7.888 billion in 2024 and shows that there are 4.95 billion active users of social media platforms, which is equal to 61% of the total population, but Facebook has many users that are approximately 3.05 billion and the remaining numbers for other social media platforms [4, 5]. The increase in users on social media platforms is directly proportional to the increasing number of cyberbullying instances [6]. Several cyberbullying detection approaches were proposed with the aim of detecting, preventing and monitoring cyberbullying instances on social media platforms to overcome the impacts of cyberbullying, but these approaches have faced various challenges in detecting cyberbullying instances [7] as follows:

(1) Cyberbullying involves several forms, such as multimodal and aggregation; therefore, many cyberbullying detection algorithms are developed to classify cyberbullying instances in text-based only, which cannot classify all forms of cyberbullying. To detect cyberbullying instances across all forms, powerful algorithms are required to investigate and understand cyberbullying instances in all modalities [8].

(2) Mostly of cyberbullying detection algorithms are developed to capture cyberbullying instances for high-resource languages like English, Chinese, French, and German, so it causes bias in detecting cyberbullying instances on low-resource languages like Swahili, Bengali, and Marathi which may affect the accuracy, precision and fairness of cyberbullying classification processes [9].

(3) Collection and labeling of data are costly and time-consuming due to restrictions of data sources platforms, which are social media platforms such as Twitter, Facebook, YouTube, and Instagram. Only the Twitter platform allows researchers to clone unlabeled data via the Twitter API for academic use without more restrictions compared to other social media platforms with tight restrictions to get data from them. High-quality training and testing data are required to train and test cyberbullying detection models to obtain effective algorithms to detect and identify cyberbullying instances on social media platforms [10].

(4) Social media platforms are changing in terms of data availability, features, and user interfaces. It is difficult to develop automated cyberbullying detection systems that are multipurpose to adapt to different digital platforms [11].

Apart from that, there is continuous development of cyberbullying strategies that produce obfuscated words such as misspellings and coded languages that aim to harm individuals, but it is difficult to identify by cyberbullying automatic detection systems, so powerful tools are required to detect these kinds of cyberbullying instances.

### 1.1 Review Taxonomy

The review covered four main approaches to cyberbullying detection on social media platforms such as Machine Learning, Deep Learning, Traditional models, and Large Language Models (LLMs). For Machine Learning, various algorithms like ANN, SVM, KNN, NB, were compared. Deep Learning explored algorithms such as DBN, DNN, CNN, and RNN. Traditional models focused on rule-based and character percentage matching techniques. LLMs, including GPT and BERT, were also examined for their effectiveness in cyberbullying detection, as illustrated in Figure 1.

### 1.2 Contributions

The contributions of the paper are outlined as follows:

(1) To systematically conduct a comprehensive comparative study of existing research on methods used to detect cyberbullying on social media platforms from 2018 to August 20, 2025.
(2) To analyze the existing technologies and approaches employed in detecting cyberbullying on social media platforms.
(3) To analyze existing evaluation metrics used in cyberbullying detection on social media platforms.
(4) To analyze the source, diversity, generalizability, size, scope, and annotation quality of datasets employed in cyberbullying detection on social media platforms.
(5) To explore existing challenges and propose pathways for future research in cyberbullying detection on social media platforms.

### 1.3 Paper Organization

The remainder of this paper is organized as follows: Section 2 outlines the review procedure; Section 3 summarizes existing surveys on cyberbullying detection; Section 4 provides background information; Section 5 reviews cyberbullying detection approaches; Section 6 discusses challenges and future directions; and Section 7 concludes the paper.

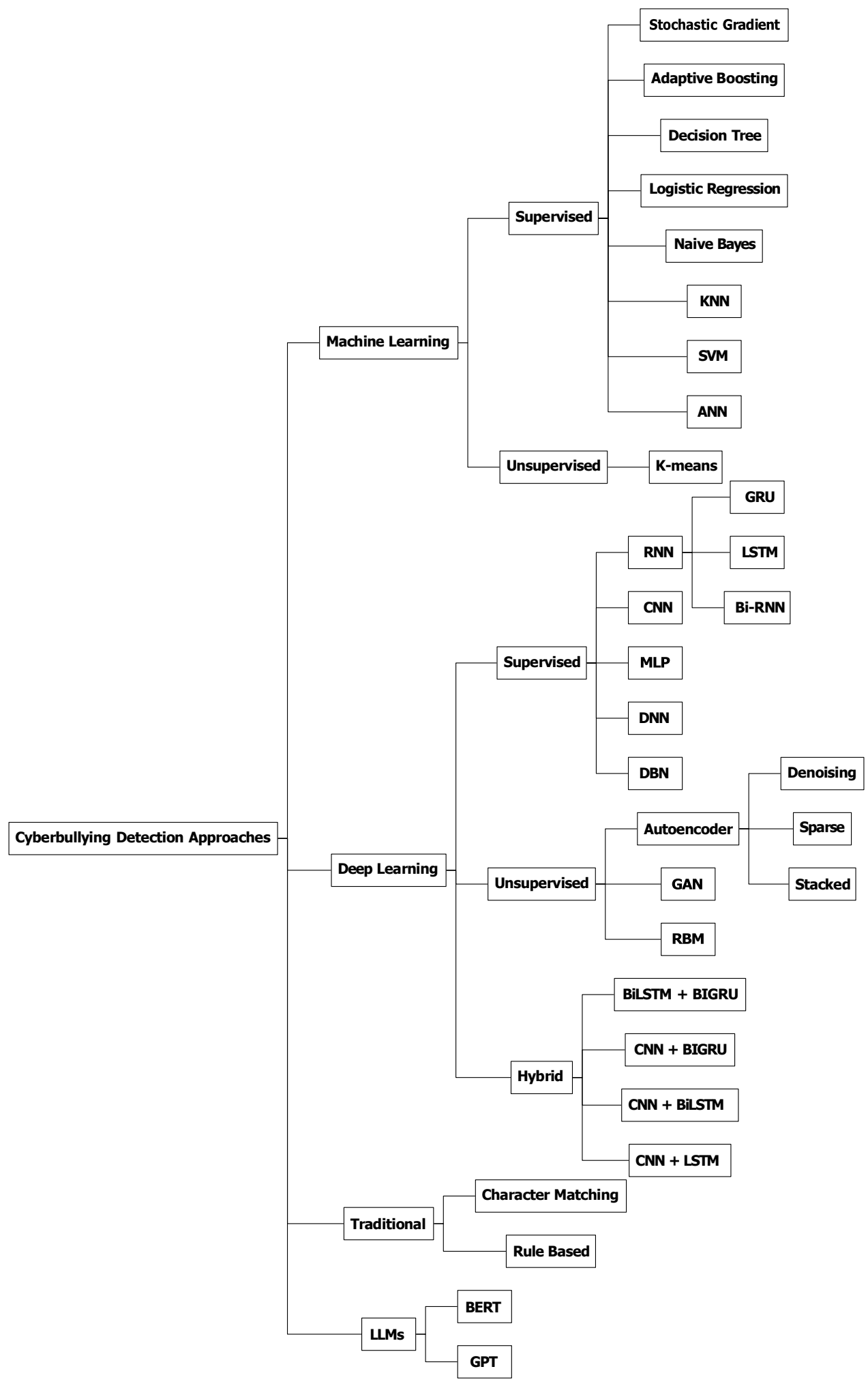

Fig. 1. Taxonomy of cyberbullying detection approaches on social media platforms.

## 2 PROCEDURE FOR CONDUCTING REVIEW

This section outlines the complete process of conducting a comprehensive systematic review following the PRISMA 2020 guidelines [12]. The methodology ensures transparency, reproducibility, and minimization of bias through clearly defined protocols for study selection, data extraction, and quality assessment.

### 2.1 PRISMA Framework Implementation

The review strictly adheres to the PRISMA framework. Figure 2 presents the PRISMA flow chart detailing the four-phase process: identification, screening, eligibility, and inclusion.

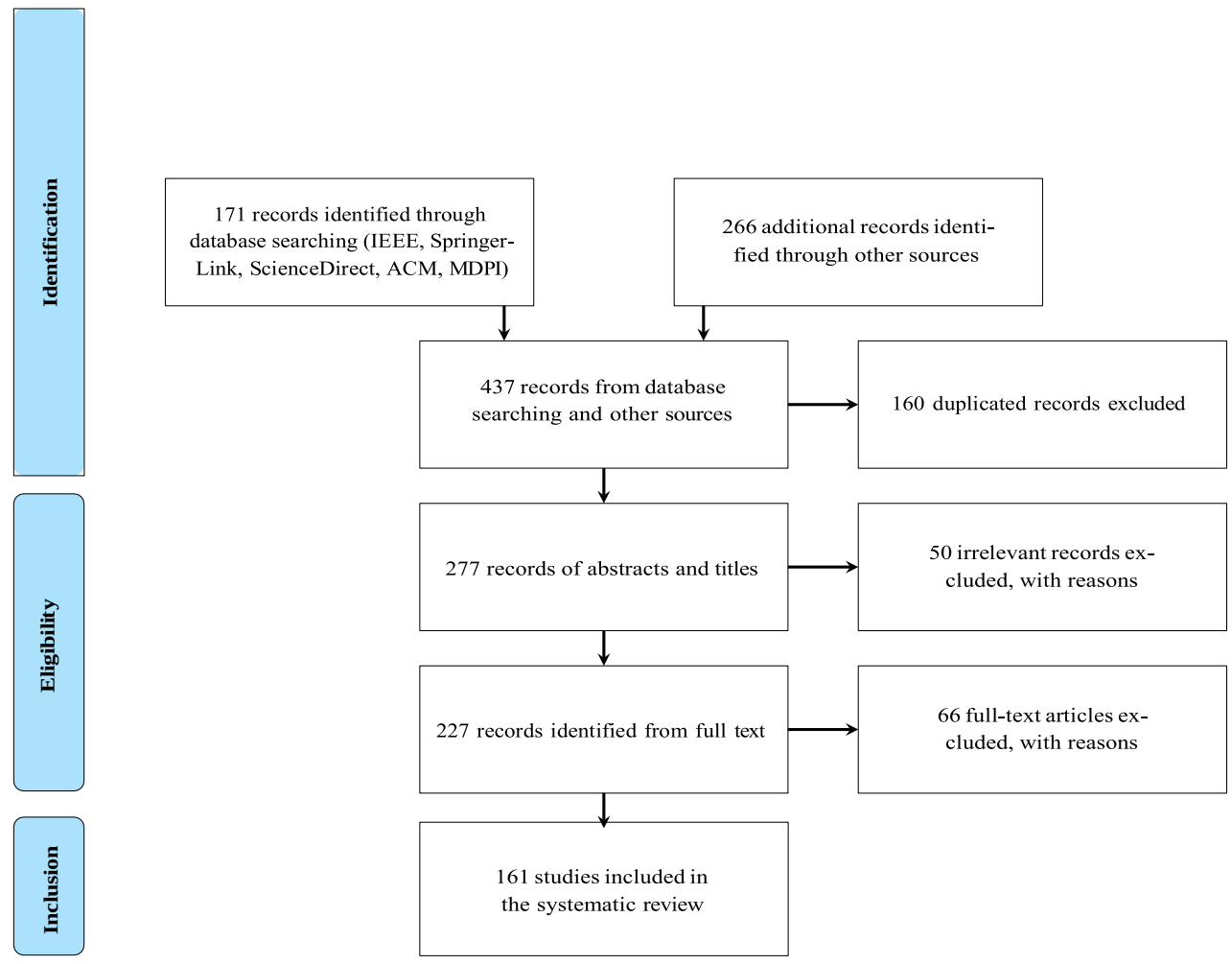

Fig. 2. PRISMA flowchart illustrating the stages of selecting research articles for this review.

## 2.2 Process of selection

*2.2.1 Search Databases.* The articles were obtained from academic scientific databases such as IEEE Explore Digital Library, SpringerLink, MDPI, ScienceDirect, ACM Digital Library, and other sources.

*2.2.2 Search Strategy and Keywords.* The search strategy combined controlled vocabulary and free text terms, using Boolean operators:

(1) Cyberbullying terms: "harassment", "cyberstalking", "hate", "abuse", "insult", "toxic", "gender", "sexism", "racism".
(2) Language terms: "high resource", "low resource".
(3) Methodological terms: "framework", "case study", "best practice".

Example query: ("high resource" OR "low resource") AND ("framework" OR "case study") AND ("harassment" OR "hate"). The searches were carried out between 2018 and November 10, 2025.

*2.2.3 Searching procedure.* In addition to the studies that correspond to the keyword pairs generated earlier, we also included several additional studies referenced in relevant reviews identified during the search process.

*2.2.4 Eligibility Criteria.*

*Inclusion.*

- Peer-reviewed studies (from 2018 to 10 November 2025) addressing cyberbullying from technical, policy, or ethical perspectives.
- Studies that present empirical, theoretical, or conceptual frameworks.
- Full-text availability in English.

*Exclusion.*

- Non-peer-reviewed or workshop papers under 4 pages.
- Studies without a broader context of cyberbullying.
- Articles that do not address cyberbullying directly.

*2.2.5 Screening and Selection.* A two-stage screening process was used.

- Stage 1: Screening of the title and abstract (n = 277).
- Stage 2: Review of the complete text (n = 227).

All screening was performed independently by two reviewers, and disagreements were resolved by consensus or by a third reviewer. A total of 161 papers were included.

## 2.3 Data Extraction and Quality Assessment

*2.3.1 Data Extraction.* A structured data extraction form was used to collect the following:

- Bibliographic details (authors, year, source).
- Type and methodology of the study (empirical, theoretical, mixed).
- Cyberbullying focus (technical, social, economic).
- Metrics used (accuracy, precision, recall, F1 score).
- Reported challenges, recommendations, and future research directions.

The extraction was carried out by two reviewers to ensure consistency.

*2.3.2 Quality Assessment.* The quality of the study was assessed on the following:

- Clarity of objectives and scope.
- Transparency in methodology.
- Relevance for cyberbullying detection.
- Replicability of results.
- Presence of bias or conflict of interest disclosures.

## 2.4 Included Studies Analysis

The final 161 studies were analyzed using the following:

(1) Quantitative synthesis of the performance benchmark metrics reported.
(2) Thematic analysis of cyberbullying detection dimensions and challenges.
(3) Trend analysis in publications years and domains (see Table 1).

Table 1. Summary of paper statistics from various sources

| **Academic Library** | **Number of Papers** |
|---|---|
| IEEE Xplore Digital Library | 88 |
| ScienceDirect (Elsevier) | 21 |
| SpringerLink | 22 |
| ACM Digital Library | 31 |
| MDPI | 9 |
| Other sources | 266 |
| **Total** | **437** |

## 2.5 Limitations

The following limitations are acknowledged:

(1) Potential language bias due to the inclusion of only English-language publications.
(2) The dynamic nature of the cyberbullying detection field may lead to the omission of very recent studies.
(3) Inconsistencies in database indexing might have led to missed studies.

## 3 RELATED SURVEY

This section reviews existing surveys on cyberbullying detection on social media platforms, highlighting their focus, methodologies, strengths, and limitations. The following are the existing surveys on cyberbullying detection on social media platforms:

Shrivastav et al. [13] presented cyberbullying detection techniques in text-based and multimodal approaches, highlighting machine learning, deep learning, visualization, and data privacy. Although the study offers a broad and useful overview, it is largely descriptive, lacks a critical analysis of methodological challenges, and does not clearly differentiate itself from related works. Its main contribution lies in consolidating computational strategies and drawing attention to ethical and interpretability issues that are often overlooked in the cyberbullying detection domain.

Anirudh Raj et al. [14] surveyed cyberbullying detection systems, covering methods such as ML, NLP, social networks, and sentiment analysis, along with data features such as text, user behavior, and interactions. The study highlights challenges, including evolving cyberbullying tactics, contextual complexity, and ethical concerns, but remains largely descriptive with limited critical evaluation. Its main contribution lies in consolidating diverse approaches and emphasizing ethical and contextual considerations that are often overlooked in the field of cyberbullying.

Hasan et al. [15] reviewed deep learning-based cyberbullying detection systems, covering models, frameworks, datasets, and data representation techniques while highlighting advantages over traditional methods. The study identifies challenges and future research directions, but lacks a detailed comparative evaluation and a limited discussion of ethical issues. Its main contribution is to consolidate state-of-the-art deep learning approaches and to emphasize key technical and practical challenges.

Elsafoury et al. [16] surveyed automated cyberbullying detection, covering feature-based and deep learning methods, data sources, and evaluation metrics, while highlighting challenges like inconsistent definitions, data imbalance, and real-time detection needs. The study provides a clear categorization and overview, but lacks comparative performance analysis and in-depth discussion of ethical issues. Its main contribution is to consolidate the methodological landscape and identify key technical and practical challenges.

Salawu et al. [17] reviewed automated cyberbullying detection methods, categorizing them into supervised learning, lexicon-based, rule-based and mixed approaches, and highlighting challenges such as limited labeled datasets and holistic cyberbullying considerations. The study provides clear classification and guidance for future research, but lacks comparative performance analysis and in-depth discussion of ethical issues.

Alkasassbeh et al. [18] compared multiple deep learning models for the detection of cyberbullying in diverse datasets, providing insight into their relative performance. The study lacks discussion on ethical issues, real-world testing, and comparison with traditional methods, but it contributes by consolidating deep learning performance findings in the domain.

Maragathavalli et al. [19] surveyed ensemble-based deep learning models for the detection of cyberbullying, focusing on multimodal data integration to improve performance. The study highlights challenges such as data imbalance and real-time detection needs, but lacks comparative evaluation, ethical discussion, and real-world applicability. Its main contribution is to consolidate ensemble-based approaches and promote holistic, multimodal predictive models.

Farag et al. [20] surveyed unsupervised techniques for cyberbullying detection, focusing on textual methods such as clustering and topic modeling. The study identifies research gaps and potential directions, but lacks performance evaluation, ethical discussion, and real-world application examples. Its main contribution is to highlight alternative approaches beyond supervised methods.

Batani et al. [21] reviewed deep learning models for cyberbullying detection, including CNN, LSTM, Bi-LSTM, RNN, and BiGRU, highlighting their effectiveness in various forms of cyberbullying and emphasizing psychological impacts. The study advocates for multimodal approaches, but lacks comparative evaluation, ethical discussion, and real-world application examples. Its main contribution is to consolidate deep learning methods and promote holistic detection strategies.

Mishra et al. [22] surveyed text analysis techniques for cyberbullying detection on social media and messaging platforms, reviewing multilingual datasets, preprocessing methods, feature representations and machine learning and neural network approaches. The study highlights research gaps and challenges, but lacks ethical discussion, comparative evaluation, and real-world examples. Its main contribution is to consolidate state-of-the-art textual cyberbullying detection methods.

Table 2. Comparative summary of cyberbullying detection surveys.

| Study | Coverage Area | | | | | |
|---|---|---|---|---|---|---|
| | Datasets | Evaluation Metrics | Ethical and Privacy | Visualization | Challenges | Future Directions |
| [13] | ✓ | ✗ | ✓ | ✓ | ✓ | ✓ |
| [14] | ✓ | ✗ | ✓ | ✗ | ✓ | ✓ |
| [15] | ✓ | ✗ | ✗ | ✓ | ✓ | ✓ |
| [16] | ✓ | ✗ | ✗ | ✗ | ✓ | ✓ |
| [17] | ✓ | ✗ | ✗ | ✗ | ✓ | ✓ |
| [18] | ✓ | ✓ | ✗ | ✗ | ✓ | ✓ |
| [19] | ✓ | ✗ | ✗ | ✗ | ✓ | ✓ |
| [20] | ✓ | ✗ | ✗ | ✗ | ✓ | ✓ |
| [21] | ✓ | ✗ | ✗ | ✗ | ✓ | ✓ |
| [22] | ✓ | ✗ | ✗ | ✗ | ✓ | ✓ |
| **Our Study** | ✓ | ✓ | ✓ | ✓ | ✓ | ✓ |

**Key:**
✗ indicates areas that are not covered; ✓ indicates areas that are successfully covered.

Table 2 summarizes key cyberbullying detection surveys, highlighting coverage areas such as dataset analysis, evaluation metrics, ethical and privacy considerations, visualization, challenges, and future directions. This study addresses the gaps not covered by existing surveys by examining dataset analysis, evaluation metrics, ethical and privacy considerations, visualization, technologies, approaches, and limitations, while providing pathways to combat cyberbullying and enhance online safety, ethics, and social responsibility.

# 4 BACKGROUND INFORMATION ON CYBERBULLYING

## 4.1 Overview

Cyberbullying, characterized by repeated and aggressive digital behavior, occurs on various platforms such as social networks, games, and messaging, targeting individuals using digital devices such as mobile phones and computers [23, 24]. Social networks connect users around the world, enabling information sharing through posts and comments [25, 26]. Specifically affecting young people, cyberbullying differs from traditional bullying in its online nature, occurring through digital communication channels rather than face-to-face interactions [27]. The COVID-19 pandemic's shift to virtual interactions intensified cyberbullying instances, especially among adolescents aged 9 to 17 [28–30]. Despite social media platforms' algorithmic efforts, cyberbullying reports have increased by nearly 70% [31].

## 4.2 Characteristics

Cyberbullying encompasses various forms, classified according to technological changes, internet usage, and communication platforms [27, 32–34]. These include abusive behavior, hate speech, online harassment, impersonation, cyberstalking, flaming, outing, doxing, denigration, exclusion, and trolling [35–45].

## 4.3 Main Reasons

Cyberbullying can stem from a variety of social, cultural, and personal factors such as appearance, academic performance, sexuality, financial status, religion, political views, cultural background, and entertainment preferences [46–56]. These diverse factors reflect the complexity of online social dynamics, where differences in identity, beliefs, or interests can trigger harmful interactions. However, due to the ever-evolving nature of online communication, new forms and motivations for cyberbullying continue to emerge, making any categorization necessarily adaptive and incomplete.

## 4.4 Impacts

Cyberbullying has profound negative effects on individuals and society, affecting them economically, socially, and psychologically. Adolescents, in particular, are highly vulnerable to these effects [57, 58]. The impacts of cyberbullying include abusing alcohol or drugs, developing eating disorders, withdrawing from social media platforms, engaging in self-harm, deleting social media profiles, experiencing suicidal thoughts, battling depression, and dealing with social anxiety [59–68]. These effects are depicted in Figure 3.

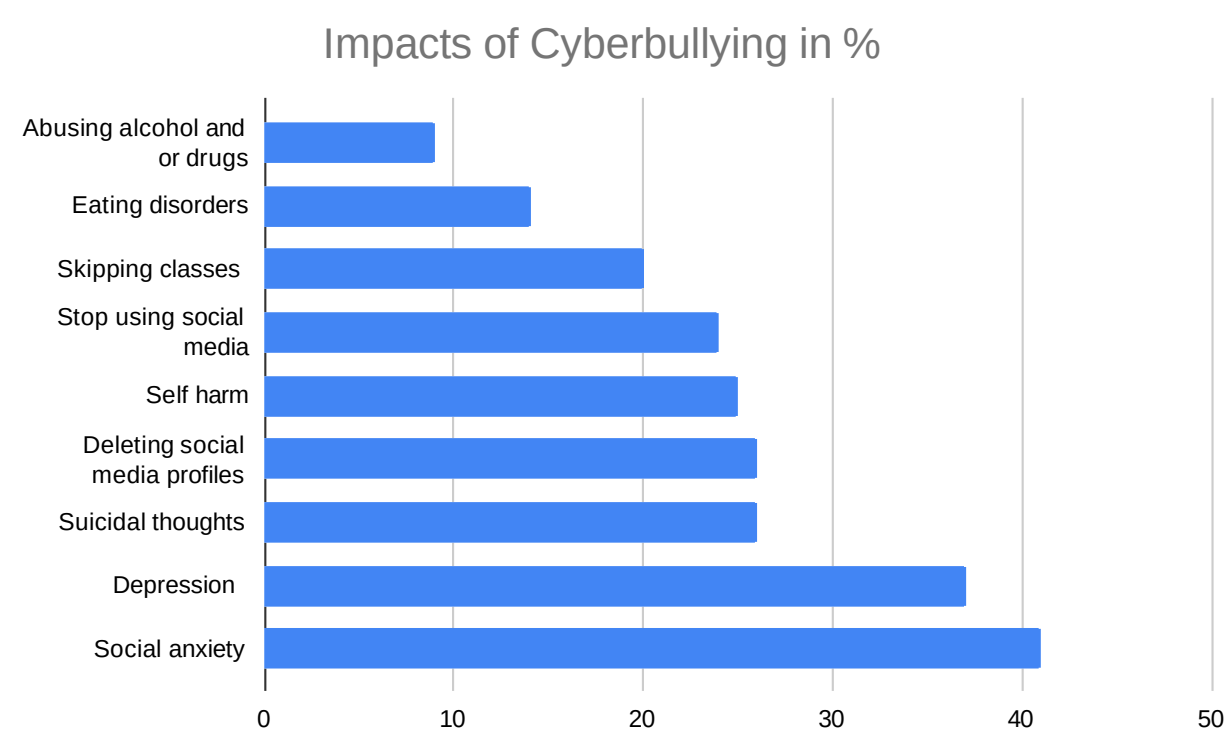

Fig. 3. Impacts of cyberbullying on individuals and communities.

# 5 REVIEW OF CYBERBULLYING DETECTION APPROACHES

This section presents a review of studies on cyberbullying detection approaches and technologies on social media platforms such as machine learning, deep learning, traditional, and large language models (LLMs). The results demonstrate the effectiveness of various linguistic features in detecting cyberbullying. Technologies, approaches, and datasets for cyberbullying detection are analyzed. These categories are illustrated in Figure 1.

## 5.1 Procedure

The cyberbullying detection methodology is divided into four main steps: data preparation, data processing, feature extraction and development, and model learning, followed by cyberbullying detection [69–71] as illustrated in Figure 4.

*5.1.1 Data Preparation.* Cyberbullying detection algorithms require data to process in order to detect and identify cyberbullying instances on social media platforms. Dataset preparation is a key step in the cyberbullying detection process, including data source, data acquisition, and data distribution [72, 73].

*5.1.2 Data Preprocessing.* Data preprocessing is a key stage in preparing data in the format required to be used as input to algorithms to detect instances of cyberbullying on social media platforms. Data preprocessing includes data cleaning, tokenization, stop words removal, normalization, data stemming, and data labeling.

*5.1.3 Feature Extraction and Development.* After collecting data from a data source, the next step is to extract features. Feature extraction is divided into two categories: extraction of feature tools such as TF-IDF, BoW, PoS tag, semantic, topic modeling, principal component analysis (PCA), linear discriminant analysis (LDA), AutoEncoder, and sentiment; and word embedding tools such as BERT, Word2Vec, GloVe, Elmo, FastText, and Farasa. Feature extraction is an important factor, since it can affect the performance of algorithms in detecting cyberbullying instances on social media platforms [74–76].

*5.1.4 Model Learning.* Model learning involves training algorithms to classify cyberbullying instances on social media platforms. These algorithms are divided into several approaches, including machine learning, deep learning, and traditional approaches. The algorithms learned after processing input data enable them to detect and identify cyberbullying instances. After training the algorithms, the next step is to test their performance in classifying cyberbullying instances using evaluation metrics such as accuracy, precision, recall, and F1 score.

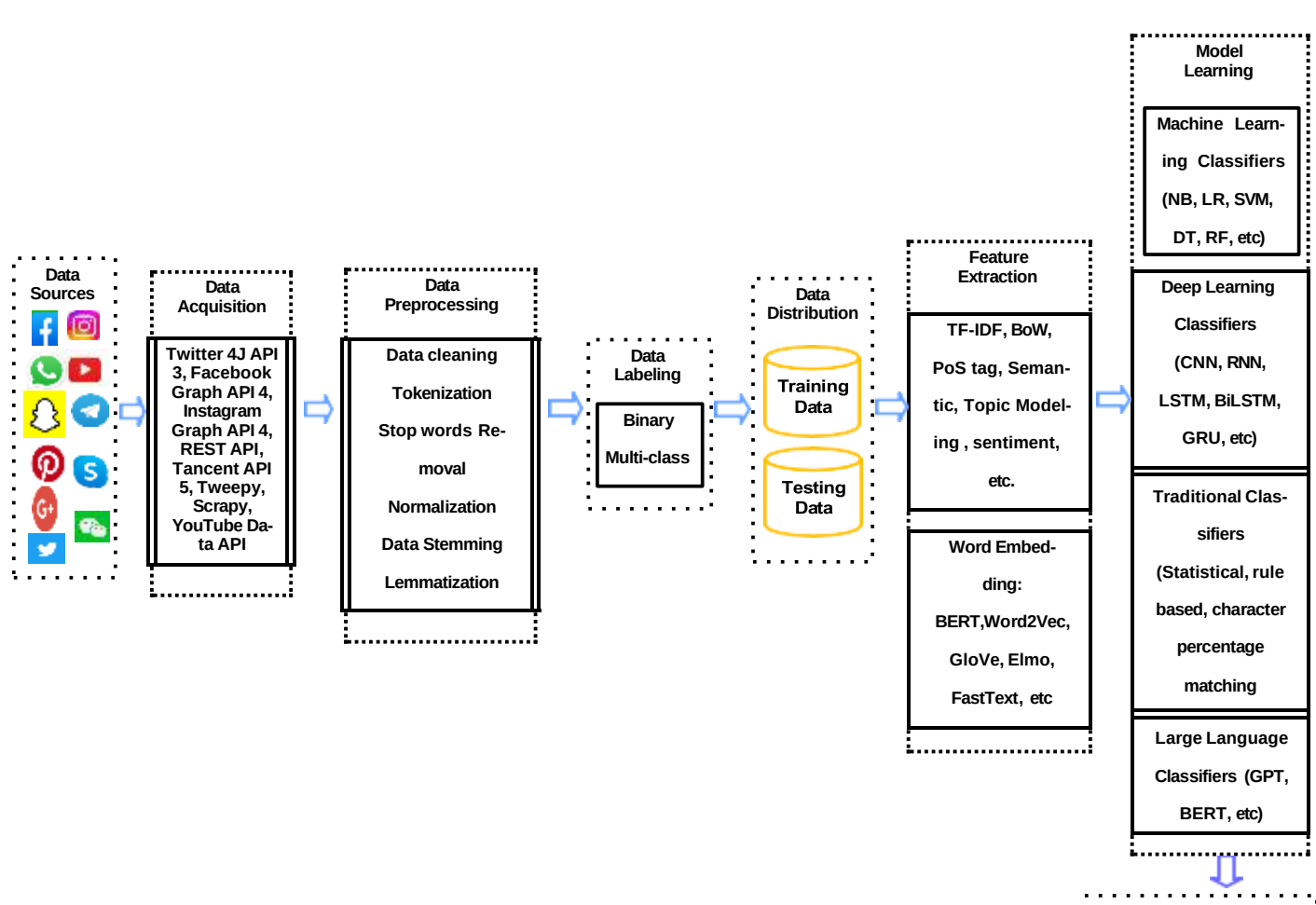

Fig. 4. A cyberbullying detection framework for social media platforms.

### 5.2 Machine Learning

This section analyzes and compares machine learning approaches for cyberbullying detection, evaluating algorithms based on accuracy, precision, recall, and F1 score. The datasets used in these studies are detailed in Table 3, and Table 4 provides an overview of existing studies, including feature extraction, classifiers, and evaluation metrics.

Table 3. Datasets used in machine learning models for cyberbullying detection on social media platforms.

| Dataset | Language | Focus | Size | Ref |
|---|---|---|---|---|
| Twitter | English | Analyze performance of ensemble model | 350,000 | [6] |
| | English, Hindi and Marathi | Multilingual system | 1,431 | [58] |
| | English, Arabic | Multilingual system | 126,704 | [77] |
| | English, Igbo | Multilingual system | 50 | [78] |
| Facebook | English, Arabic | Multilingual system | 126,704 | [77] |
| | English | Robust system | 20,002 | [79] |
| YouTube | Arabic | tweets classification | 30,000 | [80] |
| ASKfm | English, Dutch | An automatic system | 192,085 | [81] |
| Formspring | English | An automatic system | 500 | [82] |
| Wikipedia | Hindi, Hinglish | A robust framework | 115,864 | [83] |

The analysis of datasets used in machine learning models for cyberbullying detection identified challenges such as limited data usage, data imbalance, inaccurate data, lack of generalization, and dataset biases, affecting performance. Twitter is the most used platform, and English is the predominant language in the datasets.

Table 4. Review of existing machine learning models on cyberbullying detection and the mean values of evaluation metrics such as Accuracy (A), Precision (P), Recall (R), and F1-score (F), reported across the respective studies, providing a fair basis for comparison.

| Approach | Feature | Classifier | A | P | R | F | Ref |
|---|---|---|---|---|---|---|---|
| Supervised | BoW | K-NN, LR, NB, DT, RF, AB, SG, LSVC, DBC, and EC | 0.77 | 0.73 | 0.94 | 0.77 | [6] |
| | | SGD, MNB, and LR | 0.97 | 0.87 | 0.87 | 0.96 | [58] |
| | TF-IDF, BoW, N-grams | Naive Bayes, SVM | 0.815 | 0.815 | 0.270 | 0.405 | [77] |
| | TF-IDF, BoW | SVM, NB | 0.957 | 0.880 | 0.690 | 0.770 | [80] |
| | BoW, N-grams, lexicon, LDA, LSI | SVM | 0.944 | 0.567 | 0.664 | 0.612 | [81] |
| | TF-IDF, N-grams, BoW, Word2Vec | LR, SVM, RF, NB, Xgboost | 0.94 | 0.94 | 0.94 | 0.94 | [84] |
| | TF-IDF, BoW | DT, RF, SVM, NB | 0.81 | 0.80 | 0.80 | 0.80 | [79] |
| | TF-IDF | Naive Bayes, SVM | 0.976 | 0.964 | 0.931 | 0.947 | [82] |
| | TF-IDF, BoW | MNB, LR, DT, LSVC, GBoost, AdB, Bagging | 0.96 | 0.96 | 0.96 | 0.96 | [85] |
| | TF-IDF, N-grams, GloVe, Fast-Text, Paragram | XG Boost, NaÃŕve Bayes, SVM, Logistic Regression, CNN, LSTM, GRU, Bi-LSTM, Bi-GRU, CNN-BiLSTM, Att-BiLSTM | 0.950 | 0.994 | 0.968 | 0.980 | [83] |
| Unsupervised | Information Gain, Chi-Square, Pearson Correlation | DDT, ANN, SVM, RF, LR | 0.935 | 0.935 | 0.935 | 0.935 | [86] |
| | Skip grams, character tri-grams | K-means, single layer averaged perceptron | 0.84 | 0.84 | 0.83 | 0.84 | [87] |

Analysis of machine learning approaches reveals that supervised methods outperform unsupervised ones in detecting cyberbullying on social media. The SVM algorithm, in particular, excels over others. Techniques such as TF-IDF, Bag-of-Words, n-grams, and GloVe are frequently used to enhance performance.

The review covers various machine learning approaches, including SVM, LR, NB, RF, DT, XGBoost, and K-means, with SVM showing the best performance in classifying cyberbullying instances. The review also identifies several technical issues affecting performance, such as data volume, time efficiency, false positives and negatives, real-time features, biases, context understanding, adaptability, dataset size, scalability, and interpretability.

### 5.3 Deep Learning

This section analyzes and compares deep learning approaches for cyberbullying detection, reviewing existing studies to assess the strengths and weaknesses of each algorithm. Deep learning algorithms are evaluated on the basis of accuracy, precision, recall, and F1 score. The datasets used are also examined and compared, with detailed information provided in Table 5. Existing studies on deep learning approaches are summarized in Table 6, including details on the approach, feature extraction, classifiers, and evaluation metrics.

Table 5. Datasets used in deep learning models for cyberbullying detection on social media platforms.

| Dataset | Language | Focus | Size | Ref |
|---|---|---|---|---|
| Twitter | Bengali | Analyze the effectiveness of the model | 6,000 | [88] |
| | English | Perform an empirical analysis | 100,000 | [89] |
| | | An Artificial Intelligence | 40,873 | [90] |
| | | Reproduce the findings of existing studies | 92,000 | [91] |
| | | A Firefly-Based Algorithm | 39,000 | [92] |
| | | Feature subset and categorization | 1,000 | [93] |
| | | Show how models can overcome classification challenges | 128,000 | [94] |
| | | Co-trained Ensembles of Embedding Models | 4,816,345 | [95] |
| | | Fuzzy Based Genetic Operators | 10,566 | [96] |
| Facebook | Bengali | Analyze the effectiveness of the model | 6,000 | [88] |
| | English | Perform an empirical analysis | 100,000 | [89] |
| Instagram | | Perform an empirical analysis | 100,000 | [89] |
| | | Co-trained Ensembles of Embedding Models | 4,816,345 | [95] |
| | | Identify linguistic features such as idioms, sarcasm, and irony | 500 | [97] |
| YouTube | Bengali | Analyze the effectiveness of the model | 6,000 | [88] |
| | English | Reproduce the findings of existing studies | 92,000 | [91] |
| ASKfm | | Co-trained Ensembles of Embedding Models | 4,816,345 | [95] |
| Myspace | | Fuzzy Based Genetic Operators | 10,566 | [96] |
| Formspring | English, Hindi, Hinglish | Assess real-time social media posts and identify harmful instances | 12,000 | [31] |
| | English | Reproduce the findings of existing studies | 92,000 | [91] |
| Wikipedia | | Show how models can overcome classification challenges | 128,000 | [94] |

The analysis of datasets used in deep learning models for cyberbullying detection identified challenges such as limited data usage, data imbalance, inaccurate data, lack of generalization, and dataset biases, affecting performance. Twitter is the most used platform, and English is the predominant language in the datasets.

Table 6. Review of existing deep learning models on cyberbullying detection and the mean values of evaluation metrics such as Accuracy (A), Precision (P), Recall (R), and F1-score, reported across the respective studies, providing a fair basis for comparison.

| Approach | Feature | Classifier | A | P | R | F | Ref |
|---|---|---|---|---|---|---|---|
| Supervised | GloVe and FastText, N-grams | MNB, SVM, LR, and XGBoost, CNN, LSTM, BLSTM, GRU | 0.98 | 0.98 | 0.98 | 0.98 | [31] |
| | Not available | Bi-LSTM, CNN | 0.827 | 0.827 | 0.850 | 0.855 | [88] |
| | | BiLSTM, GRU, LSTM, RNN | 0.82 | 0.86 | 0.91 | 0.88 | [89] |
| | random, GloVe, SSWE | CNN, LSTM, BLSTM | 0.76 | 0.73 | 0.80 | 0.76 | [91] |
| | GloVe | CNN, SVM | 0.95 | 0.93 | 0.73 | 0.95 | [92] |
| | BCO-FSS, Pearson Correlation, Chisquared, Information Gain | SSA-DBN, ANN-DRL, ANN, SVM, RF, LR, NB | 0.999 | 0.999 | 0.999 | 0.999 | [93] |
| | random, GloVe, SSWE | CNN, LSTM, BLSTM, BLSTM with attention | 0.98 | 0.92 | 0.98 | 0.95 | [94] |
| Unsupervised | TF-IDF, BoW | TFIDF-SVM, CNG-LR, MF-LR, K-CNN, Bi-LSTM, Bi-GRU, Stacked Autoencoder (AICBF-ONS) | 0.940 | 0.931 | 0.953 | 0.940 | [90] |
| | Embedding Layer | tGraph Auto-Encoder (GAE) | 0.88 | 0.70 | 0.67 | 0.962 | [97] |
| | Textual and social features | LSTM | 0.67 | 0.65 | 0.67 | 0.61 | [95] |

Analysis of deep learning approaches reveals that supervised methods outperform unsupervised ones in detecting cyberbullying on social media, and the BiLSTM algorithm excels. Techniques such as TF-IDF, random, and GloVe are frequently used to enhance performance.

The review covers various deep learning approaches, including CNN, LSTM, BiLSTM, GRU, BiGRU, and stacked Autoencoder, with BiLSTM showing the best performance in classifying cyberbullying instances. The review also identifies several technical issues that affect performance, such as data dependency, generalization, interpretability, overfitting, dataset size, adaptability, and explainability.

## 5.4 Traditional Models

This section analyzes and compares traditional approaches for cyberbullying detection, investigating various studies to assess the strengths and weaknesses of each algorithm. Performance metrics such as accuracy, precision, recall, and the F1 score are used for comparison. The datasets used in these studies are detailed in Table 7, providing information on language, focus, size, and references. In addition, Table 8 summarizes existing studies on traditional approaches, describing the method, feature extraction, classifiers, evaluation metrics, and references.

Table 7. Datasets used in traditional models for cyberbullying detection on social media platforms.

| Dataset | Language | Focus | Size | Ref |
|---|---|---|---|---|
| Ofcom | English | Classify obfuscated abusive words in English text | 200 | [98] |
| TUKI | Swahili | Classify obfuscated abusive words in Swahili text | 200 | [98] |

The analysis of datasets used in traditional cyberbullying detection models revealed challenges such as limited data usage and lack of generalization, affecting performance. TUKI and Ofcom were identified as data sources for Swahili and English, respectively.

Table 8. Review of existing traditional models on cyberbullying detection and the mean values of evaluation metrics such as Accuracy (A), Precision (P), Recall (R), and F1-score (F), reported across the respective studies, providing a fair basis for comparison.

| Approach | Feature | Classifier | A | P | R | F | Ref |
|---|---|---|---|---|---|---|---|
| Statistical | TF-IDF, N-grams, Lexicon | rule based and, character percentage matching | 0.96 | 0.71 | 0.73 | 0.97 | [98] |

Analysis of traditional models reveals that statistical approaches such as rule-based and character percentage matching techniques outperform current methods in the detection of cyberbullying on social media platforms. Techniques such as TF-IDF, n-grams, and Lexicon also improve performance.

The review shows that traditional approaches, including rule-based character percentage matching, effectively classify cyberbullying instances on social media. However, the review identifies technical issues that affect performance, such as context understanding, scalability, high false-positive rates, insensitivity to semantic meaning, and adaptability.

## 5.5 Large Language Models (LLMs)

This section analyzes and compares large language models for cyberbullying detection, such as BERT, RoBERTa, DistilBERT, and GPT-3, as shown in Figure 1. Various studies are reviewed to assess the strengths and weaknesses of each model. Performance metrics such as accuracy, precision, recall, and the F1 score are used for comparison. The datasets used in these studies are detailed in Table 9, including information on language, focus, size, and references. Table 10 summarizes existing studies on large language models, detailing the approach, feature extraction, classifiers, evaluation metrics, and references.

Table 9. Datasets used in large language models for cyberbullying detection on social media platforms.

| Dataset | Language | Focus | Size | Ref |
|---|---|---|---|---|
| Twitter | English | Cyberbullying detection without the connection to platform moderation | 47,000 | [99] |
| | | Use Large language models (LLMs) for cyberbullying detection | 99,544 | [100] |
| YouTube | | Cyberbullying detection without the connection to platform moderation | 47,000 | [99] |
| | | Use GPT-3 to identify sexist and racist text passages | 1,431 | [101] |
| | | Use GPT-3 for hate speech detection | 1,000 | [102] |
| Reddit | | Assessing online misogyny using GPT-3.5 | 6,567 | [103] |
| | | Cyberbullying detection without the connection to platform moderation | 47,000 | [99] |
| | | Use GPT-3 for hate speech detection | 1,000 | [102] |
| | | Use GPT-3 to identify sexist and racist text passages | 1,431 | [101] |
| Formspring | | Use Large language models (LLMs) for cyberbullying detection | 99,544 | [100] |
| HateXplain | | Hate speech detection | 99,544 | [100] |
| OLID | | Addressing the limitations of manual data annotation processes | 14,100 | [104] |
| HASOC | English and Bengali | Classification tasks for Hate Speech | 1,776 | [105] |

Analysis of datasets for large language models in cyberbullying detection reveals challenges such as limited data usage, data imbalance, lack of generalization, and real-time detection issues, all of which affect model performance. Reddit and YouTube datasets are used more frequently than those of other platforms, with English being the predominant language.

Table 10. Review of existing large language models on cyberbullying detection and the mean values of evaluation metrics such as Accuracy (A), Precision (P), Recall (R), and F1-score (F), reported across the respective studies, providing a fair basis for comparison.

| Approach | Feature | Classifier | A | P | R | F | Ref |
|---|---|---|---|---|---|---|---|
| Unsupervised | Self-attention | GPT-3-Ada model, BERT base, Multilingual BERT, RoBERTa, DistilBERT | 0.90 | 0.92 | 0.88 | 0.93 | [99] |
| | | BERT, GPT-3.5-turbo API | 0.91 | 0.47 | 0.63 | 0.54 | [103] |
| | | BERT, XLNet, RoBERTa, XLM - RoBERTa | 0.87 | 0.86 | 0.87 | 0.87 | [100] |
| | | GPT-3 Text-davinci-001 | 0.805 | 0.755 | 0.655 | 0.685 | [101] |
| | | BERT, RoBERTa, XLNet, DistilBERT, and GPT-2.0 | 0.95 | 0.95 | 0.95 | 0.95 | [106] |
| | | ALBERT, DeBERTa, ELECTRA, HateBERT, and DeepSeek-coder-1.3b-base | 0.9185 | 0.9185 | 0.9185 | 0.9184 | [107] |

Analysis of large-language models reveals that these unsupervised approaches excel in detecting cyberbullying on social media platforms, with BERT outperforming other algorithms. Techniques like self-attention enhance performance.

The review shows that large language models, including GPT, BERT, XLNet, and XLM, effectively classify cyberbullying instances, especially BERT. The review also identifies technical issues that affect performance, such as scalability, context understanding, and biases.

## 5.6 Strengths and Weaknesses

Cyberbullying detection approaches are reviewed to identify their potential strengths and weaknesses in classifying cyberbullying instances on social media platforms. The strengths and weaknesses of cyberbullying detection approaches are summarized in Table 11.

Table 11. Summary of strengths and weaknesses of the cyberbullying detection approaches.

| Strengths | Weaknesses | Year | Ref |
|---|---|---|---|
| Moderate accuracy,large dataset, properly handles overfitting and high-dimensional spaces | Data imbalance, adaptability, interpretability and scalability, extensive data, slower training time, false positives and negatives, and real-time detection | 2018 | [81] |
| High accuracy, large dataset, automatic feature extraction and detection, faster training time, scalability, and interpretability | Data imbalance, struggles with typos and misspellings | 2018 | [92] |
| High accuracy, large dataset, scalability, adaptability, and interpretability, automatic feature representation | Data imbalance, overfitting, and generalization | 2018 | [94] |
| High accuracy, multilingual detection, relatively quick training time, handling overfitting, scalability, interpretability | Limited dataset and feature representation, lack of sentiment and sarcasm | 2019 | [58] |
| Reasonable accuracy, proper handling of overfitting and high-dimensional spaces | Limited data, Data imbalance, adaptability, interpretability and scalability, extensive data, slower training time, false positives and negatives, and real-time detection | 2020 | [79] |
| High accuracy, scalability, and interpretability, as well as quick training time | limited and imbalance dataset | 2020 | [82] |
| High accuracy, large dataset, automatic feature extraction and detection, faster training time, scalability, and interpretability | Dataset imbalance and computational complexity | 2020 | [89] |
| High accuracy, large dataset, faster training time, handling overfitting, scalability, interpretability | It is limited to English datasets. And, also information of users with the fields of age and gender of twitter datasets are not available | 2021 | [6] |
| High accuracy, quick data training time | Limited dataset and generalization, overfitting, adaptability, scalability, interpretability, and intensive time for rules creation | 2021 | [98] |

Table 11. Summary of strengths and weaknesses of the cyberbullying detection approaches. (Continued)

| Strengths | Weaknesses | Year | Ref |
|---|---|---|---|
| High accuracy, large dataset, faster training time, handling overfitting, scalability, interpretability | Datasets imbalance and lack of Multiple diversified datasets | 2021 | [85] |
| High accuracy, large dataset, automatic feature extraction and detection, faster training time, scalability, and interpretability | Dataset imbalance | 2021 | [83] |
| High accuracy, large dataset, faster training time, automatic detection, and multi-feature representation. It reduced restricted node related with overfitting | Dataset imbalance and dimensional space problem is still alive | 2021 | [86] |
| High accuracy, faster training time, automatic feature extraction, scalability, transfer learning, adaptability, and interpretability | Limited dataset, dataset imbalance, limited control, overfitting, and explainability | 2021 | [101] |
| High accuracy, real-time detection and faster training data | Limited Datasets, Dataset imbalance, and Lack of User Graphical Interface | 2022 | [108] |
| High accuracy, multilingual detection and faster training time | Limited datasets | 2022 | [78] |
| High accuracy, relatively quick training time, handling overfitting, scalability, interpretability | Dataset imbalance, context expert knowledge, feature representation | 2022 | [84] |
| High accuracy, large datasets, real-time detection, automatic feature extraction, faster training time, interpretability, multilingual detection, global features, scalability, and long-term dependencies | Dataset imbalance | 2022 | [31] |
| High accuracy, large dataset, automatic feature extraction, faster training time, interpretability, scalability, adaptability | Dataset imbalance, and unspecified data sources and sampling of data | 2022 | [93] |
| High accuracy, large dataset, faster training time, scalability, adaptability, and interpretability | Dataset imbalance | 2022 | [90] |
| High accuracy, large dataset, faster training time, automatic feature extraction, scalability, transfer learning, adaptability, and interpretability | Dataset imbalance, training data discrepancies | 2022 | [99] |
| High accuracy, large dataset, faster training time, automatic feature extraction, scalability, transfer learning, adaptability, and interpretability | Dataset imbalance, unstable predictions, struggle in controversial posts | 2022 | [103] |
| High accuracy, large datasets, real-time detection, automatic feature extraction, faster training time, interpretability, adaptability, transfer learning, and scalability | Datasets are skewed, datasets imbalance, the length of the posts varies between datasets in terms of the number of words | 2022 | [91] |
| High accuracy, large datasets, multilingual detection, proper handling of overfitting and high-dimensional spaces | Data imbalance, adaptability, interpretability and scalability, extensive data, slower training time, false positives and negatives, and real-time detection | 2022 | [77] |
| High accuracy, large dataset, faster training time, automatic feature extraction, scalability, transfer learning, adaptability, and interpretability | Dataset imbalance, limited control, overfitting, explainability | 2022 | [102] |
| High accuracy, automatic feature extraction and detection, faster training time, scalability, interpretability, and context understanding in long and short text | Limited dataset, dataset imbalance, lack of generalization, and data scarcity | 2023 | [100] |
| High accuracy, large dataset, automatic feature extraction and detection, faster training time, scalability, and interpretability | Data imbalance, struggles with typos and misspellings, and lack of contextual information | 2023 | [88] |
| Reasonable accuracy, large datasets, proper handling of overfitting and high-dimensional spaces. | Data imbalance, adaptability, interpretability and scalability, extensive data, slower training time, false positives and negatives, and real-time detection | 2023 | [80] |

## 6 DISCUSSION, CHALLENGES, AND FUTURE DIRECTIONS ON CYBERBULLYING DETECTION APPROACHES

This section explores recent detection approaches for classifying cyberbullying texts on social media over the past five years, examining studies in different languages. Key challenges in cyberbullying detection on social media are presented and public datasets are analyzed to identify issues affecting model effectiveness. Challenges are categorized into language, dataset, and approach-related aspects, with potential future directions outlined. Finally, solutions and recommendations are proposed to detect, prevent, and monitor cyberbullying on social media platforms.

### 6.1 Discussion

*6.1.1 High-Resource Language-Based Classification.* Most existing studies on cyberbullying detection focus on high-resource languages such as English, despite the global prevalence of cyberbullying across all languages. As social media usage increases in developing regions, more users communicate in native or local languages, which has led to under-representation of low-resource languages in current research [58]. To achieve equitable and effective detection, future models must incorporate multilingual and low-resource language data to improve generalizability and inclusiveness [109].

*6.1.2 Representation of Feature Vectors.* The review shows that both machine learning and deep learning approaches rely on effective feature vector representations for detecting cyberbullying text. Traditional machine learning methods depend on explicit feature engineering, while deep learning models automatically learn meaningful representations from raw data [84].

*6.1.3 Sparse Data Representation.* Several studies adopted sparse data representation techniques to handle datasets containing many missing or zero-valued entries efficiently [110]. Such representations improve computational efficiency by storing only non-zero values and their indices [111]. Moreover, expert or linguist-assisted labeling remains critical to ensure data reliability during the annotation process.

*6.1.4 Effectiveness of Cyberbullying Detection Approaches.* Machine learning classifiers demonstrate limitations in automatic feature extraction, real-time detection, and scalability for large datasets [112–114]. Deep learning models, although superior in semantic understanding, still face real-time detection challenges [15]. Similarly, traditional classifiers struggle to capture linguistic nuances, while large language models (LLMs) lack efficient mechanisms for real-time inference and deployment [115].

*6.1.5 Linguistic Diversity in Cyberbullying Detection Datasets.* Cyberbullying detection research spans several languages, reflecting the multilingual nature of online communication are outlined as follows:

(1) *English.* Most studies employ English datasets due to the language's dominance on social media platforms and the abundance of NLP resources such as dictionaries, embeddings, and sentiment lexicons [116–120].
(2) *Swahili.* Widely spoken in East and Central Africa (e.g., Tanzania, Kenya, Uganda, Rwanda, Malawi, Zambia, South Africa), Swahili datasets provide valuable insights into regional cyberbullying patterns.
(3) *Arabic.* Used across the Middle East and North Africa, Arabic datasets help uncover linguistic and cultural dimensions of cyberbullying in these regions.
(4) *Hindi.* With India's vast online population, Hindi datasets enable localized understanding of cyberbullying behaviors.

(5) *Dutch.* Dutch datasets, though smaller, contribute to regional studies focused on European online interactions.

*6.1.6 Sources of Datasets for Cyberbullying Detection.* Datasets for cyberbullying detection are primarily collected from two main types of social media sources are outlined as follows:

(1) *Twitter Datasets.* Twitter remains the most widely used source due to its accessible APIs, permissive data-use policies, and real-time data streaming capabilities. Its large and active user base provides rich textual data, enabling timely analysis of evolving cyberbullying trends [11, 118, 121–123].
(2) *Other Platforms.* Additional data sources include Facebook, Instagram, Reddit, FormSpring, MySpace, ASKfm, Ofcom, TUKI, and Wikipedia. Each platform differs in architecture, user interaction style, and data-collection restrictions. For instance, Facebook and Instagram prioritize visual content and enforce stricter data-access policies, which complicates textual analysis. Although Twitter dominates due to accessibility and text-rich data, incorporating diverse platforms would offer a more holistic understanding of cyberbullying behaviors across modalities.

*6.1.7 Ethical Considerations.* Cyberbullying detection studies must carefully address ethical concerns due to the sensitivity of user-generated content. In line with General Data Protection Regulation (GDPR) and the California Consumer Privacy Act (CCPA) compliance, datasets should contain no sensitive or personal information, and only publicly available data should be used. Any contact information must be removed and anonymized to protect user identities [107, 124]. The deployment of cyberbullying detection systems raises important ethical concerns that must be addressed to ensure responsible and fair use [125]. The ethical considerations in cyberbullying detection on social media platforms are outlined as follows:

(1) *Privacy.* Cyberbullying detection systems often require access to personal communications, user profiles, or behavioral data [126]. Without strong safeguards, the collection and processing of such information can infringe upon users' rights to privacy and data protection [127].
(2) *Consent.* In cyberbullying detection systems, users are rarely informed or asked for approval when their online interactions are monitored by automated detection tools [127]. This lack of transparency raises concerns regarding consent, as individuals may be unaware that their private communications and behavioral patterns are being analyzed [128].
(3) *Bias and Fairness.* Cyberbullying detection systems trained on imbalanced or language-specific datasets may disproportionately misclassify certain groups, leading to unequal protection and potential stigmatization of marginalized communities [129, 130]. This risk highlights the importance of transparency in dataset construction and the need for inclusive, diverse training data [131].
(4) *Responsible use of detection.* Cyberbullying detection should not only focus on punitive measures but also prioritize supportive interventions, such as counseling or educational awareness, that safeguard victims without unjustly penalizing individuals [126]. Policymakers, educators, and technology providers must collaborate to establish ethical guidelines that balance the benefits of early detection with respect for user autonomy and rights.

## 6.2 Challenges

*6.2.1 Language.* These are challenges related to language issues in cyberbullying detection on social media platforms as follows:

(1) *Diverse Linguistic Structure.* Social media platform messages often feature informally written text that may not adhere to formal linguistic structures, making it challenging to distinguish patterns within the text. Cyberbullying encompasses a variety of linguistic forms, including nonverbal modes of expression, code mixing, and the use of dissimilar languages. The rise of social media platforms and communication in digital systems has introduced novel avenues for cyberbullying [132]. Factors such as the utilization of typographic elements like capitalization, punctuation, and emojis present additional complexities in identifying cyberbullying within social media platform instances [133]. Moreover, the presence of cultural disparities and region-specific topics on social media platforms increases the diversity and quantity of user-generated instances, thereby complicating the linguistic dimensions of cyberbullying detection [134]. Considering these linguistic dynamics is vital for devising effective strategies to identify and mitigate cyberbullying in online forums [135]. Researchers endeavor to dissect language patterns and multimodal forms of expression to gain deeper insights into cyberbullying and its various manifestations across linguistic landscapes [136].

(2) *Cultural variations.* Cyberbullying is often subjective and contingent on context. What may be perceived as cyberbullying in one context might not be considered as such in another. Accurately detecting cyberbullying requires a comprehensive understanding of context and cultural intricacies, posing a challenge for algorithms. For example, in African countries, referring to a person from Europe as "European" may be seen as a sign of respect or wealth, while calling a person from Africa "African" may be viewed as racism. Understanding cyberbullying involves recognizing the influence of cultural elements. Various cultural factors, including gender, race, ethnicity, disability, and sexual orientation, shape the dynamics of cyberbullying within workplace settings [137]. The Barlett Gentile cyberbullying model (BGCM) has been identified as a robust framework for predicting cyberbullying behaviors and attitudes across diverse cultural contexts [138]. Studies indicate that cyberbullying is a global phenomenon, with societal and cross-national differences in rates of bullying and victimization, as well as in the characteristics of cyberbullying instances [139, 140]. These disparities can be attributed to cultural norms, educational systems, technological infrastructure, legal frameworks, and socio-economic disparities [141]. Cyberbullying research has also focused on understanding the experiences of various marginalized youth populations, underscoring the necessity of acknowledging cultural diversity in comprehending and combating cyberbullying.

(3) *Context Misinterpretation.* The interpretation of cyberbullying heavily relies on context. Without a comprehensive understanding of the context in which these statements are made, distinguishing between cyberbullying and non-cyberbullying statements can be challenging. Misinterpreting context may lead to varying findings and overstated claims regarding the prevalence, progression over time, and impacts of cyberbullying [142]. It may result in cyberbullying becoming conflated with general cyber harassment, complicating efforts to measure and comprehend the exact phenomenon [143]. Misinterpreting context may also delay efforts to effectively tackle and stop cyberbullying, as it might not be recognized as a distinct characteristic of cyberbullying [144]. Furthermore, a lack of comprehension of the context surrounding cyberbullying can hinder the creation of precise and well-organized tools for identifying risks, as external viewpoints may not fully grasp the experiences and perspectives of those involved [145]. In summary, a deficiency in understanding the context of cyberbullying can impede advancements in research, prevention, and intervention activities.

*6.2.2 Datasets.* These are challenges related to labeling and annotation issues in cyberbullying detection on social media platforms, as follows:

(1) *Imbalance of the dataset classes.* In many studies, an imbalance of the dataset classes is identified in many studies. The imbalance of data has a significant impact on the training and performance of the models. Class imbalance occurs when either the cyberbullying or non-cyberbullying class is more prevalent than the other class, often defined by majority to minority class ratios. The imbalance of datasets is categorized into three groups: low-class, moderate-class, and high-class imbalance. The low or accepted and moderate class imbalances are in a 10: 1 ratio, but the high class imbalance ranges between 100:1 and 10000:1 [146], as shown in Table 12. The review identified that non-cyberbully class instances were more prevalent than bullying class instances. This imbalance led to several challenges in model training and evaluation. Firstly, bias in model training: this led models to become more motivated to detect non-cyberbullying than the cyberbullying class and tended to label most instances as non-cyberbullying [94]. Secondly, poor generalization: this led the model to struggle in generalizing new and unseen data in the cyberbullying class because the model had not seen enough examples of the cyberbullying class during training to learn robust patterns [91]. Thirdly, misleading evaluation metrics: models predicting the non-cyberbullying class for every instance might still achieve high accuracy, but accuracy alone is insufficient for evaluating the true performance of the model in imbalanced classes. Therefore, other evaluation metrics such as precision, recall, and the F1 score are needed to evaluate the true performance of the model. Furthermore, it was difficult to set a threshold for detecting cyberbullying due to bias towards the non-cyberbullying class [99].

Table 12. Imbalance of the dataset classes in Cyberbullying Detection on Social Media Platforms.

| **Dataset** | **Class** | **Count** | **Percentage** | **Ratio** | **Remark** | **Ref** |
|---|---|---|---|---|---|---|
| Twitter | sexist | 3,117 | 19.5% | 10000:1 | high class imbalance | [94] |
| | racist | 1,937 | 12.1% | | | |
| | neither sexist nor racist | 10,000 | 68.4% | | | |
| | offensive | 4,929 | 54.2% | 10:1 | low class imbalance | [85] |
| | non-offensive | 4,164 | 45.8% | | | |
| FormSpring | cyberbullying | 776 | 6.1% | 10000:1 | high class imbalance | [100] |
| | non-cyberbullying | 11,997 | 93.9% | | | |
| Facebook | cyberbullying | 2,196 | 6.2% | | | [77] |
| | non-cyberbullying | 33,077 | 93.8% | | | |
| Wikipedia | cyberbullying | 13,590 | 11.7% | | | [83] |
| | non-cyberbullying | 102,274 | 88.3% | | | |
| Reddit | misogynistic | 129 | 9.9% | | | [103] |
| | non-misogynistic | 1,174 | 90.1% | | | |
| ETHOS | hate speech | 433 | 43.4% | 10:1 | low class imbalance | [101] |
| | non hate speech | 565 | 56.6% | | | |

(2) *Limitation of Dataset.* The dataset may be viewed as limited due to factors such as incomplete data percentages, inadequate details for comprehensive analytical evaluations, and the utilization of a small dataset for training learning models [110]. Based on the utilization of a small dataset for training learning models to detect cyberbullying on social media platforms, the review identified that a small volume of data was used to train models for the classification of both cyberbullying and non-cyberbullying text [147, 148]. The results of the models showed that they had uninterested performance, less than a performance ratio of 0.5, since they were trained with a small volume of data and did not have enough samples of instances of cyberbullying or non-cyberbullying to learn strong patterns [80, 111]. The classification of a large dataset significantly improves the performance of the models in classifying cyberbullying or non-cyberbullying texts [78, 149].

(3) *Inconsistent Annotations.* Labeling of datasets requires experts and linguists who have a comprehensive understanding of the language context to produce accurate datasets. This leads to a shortage of labeled social media platform datasets suitable for training effective cyberbullying detection systems [150]. Sometimes errors can occur during data annotations, leading to data labeling inconsistency. These errors may be caused by various reasons, such as subjective labeling, ambiguous definitions, cultural and contextual variations, inter-annotator agreement, and errors in the annotation process [151]. In response to this challenge, automated methods, such as self-supervised learning (SSL) models, have been suggested to label data and enrich training sets using unlabeled data [111]. Furthermore, a study introduces a labeled Instagram dataset with comprehensive annotations on critical cyberbullying characteristics, offering fresh perspectives for automated detection [152]. A survey paper discusses the need for improved and more precise detection techniques, along with the exploration of alternative annotation methods [153].

(4) *Extremely Inaccurate data.* Data may be considered inaccurate due to significant deviations from the actual values. The review identified the presence of extremely manipulated data in existing studies on cyberbullying text detection, which deviated significantly from actual data from social media platforms used to train and evaluate the performance of the models. The impact of using inaccurate data is evident when it comes to the real situation of cyberbullying detection on social media platforms, as models perform poorly due to the misclassification of some cyberbullying instances. Access restrictions on high-quality data constrain the utility of cutting-edge techniques, thereby limiting their applicability [114]. Small datasets impede direct comparison of progress and assessment of applicability, as they do not adequately capture complex social dynamics [110]. Using incorrect data may result in decreased classifier performance, as it overlooks the nuances and complications present in real-life bullying situations [154].

(5) *Single-label Based Classification.* The review identified that many studies conduct research on cyberbullying detection on social media platforms based on single labels or binary labels, which categorize instances as either cyberbullying or non-cyberbullying. However, technology evolves rapidly and bullies continuously develop new methods to harass others. In addition, there are various forms of cyberbullying, including sarcasm, toxicity, hate speech, and attacks based on ethnicity and religion, which are used to target individuals on social media platforms [32]. Therefore, data labeling should not be limited to distinguishing between cyberbullying and non-cyberbullying but should also include other forms of cyberbullying to increase the effectiveness of the model in classifying cyberbullying texts [27].

*6.2.3 Approaches.* These are challenges related to approach issues in cyberbullying detection on social media platforms, as follows.

(1) *Limited Ability of Multilingual Pre-Trained Language Models.* Expanding cyberbullying detection to various languages presents unique challenges specific to each language. These challenges may include differences in linguistic structures and cultural expressions. Detecting cyberbullying presents a challenge for multilingual pre-trained language models due to their struggle with extended social media platform sessions and the issue of imbalanced classes. Although models like BERT and RoBERTa excel in natural language processing tasks, they have not been widely utilized for the detection of cyberbullying [150]. The prolonged duration of social media platform interactions poses a difficulty for transformer models, which have input length constraints. A solution to this challenge is LS-CB, a framework that combines predictions from transformer models across smaller sliding windows, leading to improved performance [100]. Moreover, the scarcity of quality cyberbullying datasets in languages with limited resources obstructs the development of effective detection models for local contexts [155]. Transfer learning, particularly fine-tuning pre-trained language models such as DistilBERT and ELECTRA-small, has demonstrated promising outcomes in cyberbullying detection, surpassing traditional machine learning methods [156].

(2) *Limited Model Interpretability.* In cyberbullying detection, the lack of a clear understanding of how models make predictions presents a significant challenge [150]. The use of cutting-edge methods is restricted by restricted access to top-tier data, leading to the creation of small, varied datasets that impede straightforward comparisons of advancements [110]. Current cyberbullying detection models are incomplete or demonstrate restricted effectiveness when applied to real-time situations [157]. Machine learning, deep learning, traditional, and large language models have shown that they can detect instances of cyberbullying on social media platforms, but they lack transparency in the decision-making process. This makes it difficult for humans to understand what is done behind the scenes and what causes the performance of the model to be either low, moderate, or high [158].

(3) *Transfer Learning and Fine-Tuning Challenges.* Utilizing transfer learning and fine-tuning techniques demands extensive computational resources and time, along with abundant data to enhance the detection capabilities of models [155]. The effectiveness of these approaches heavily relies on the quality and suitability of the pre-trained models employed. Should the pre-trained models not align well with the specific requirements of cyberbullying detection, the performance might suffer [159]. Furthermore, addressing linguistic complexities inherent in user-generated instances across various languages poses a challenge for transfer learning and fine-tuning. These complexities involve cultural nuances, unconventional use of linguistic resources, and the availability of native language input methods [160]. Hence, while transfer learning and fine-tuning exhibit potential in cyberbullying detection, they come with limitations and may not always represent the most optimal strategy.

(4) *Lack of Real-Time Detection.* Various classifiers may detect cyberbullying instances in different forms on social media platforms. As users generate diverse cyberbullying instances continuously, classifiers often lack real-time detection capabilities. This limitation stems from the rapid generation of various forms of cyberbullying instances on social media platforms [161]. Additionally, classifiers may struggle to interpret nuances in human-based expressions, resulting in failures to identify instances of bullying [15].

## 6.3 Future Direction

The future direction for advancing cyberbullying detection on social media platforms has been categorized into three aspects, namely: Datasets, languages, impacts, detection, defense, and mitigation.

*6.3.1 Datasets.* These are future studies related to dataset issues in cyberbullying detection on social media platforms, as follows:

(1) *Generation of Datasets.* Data will be collected from social media platforms and news websites, then annotated to ensure the dataset is accurate and balanced, especially for languages with low resources like Swahili. To mitigate dataset bias, it is essential that these datasets cover a wide range of languages, dialects, cultural backgrounds, and various forms of offensive language. Employing a broader range of strategies for data collection, such as sampling from multiple platforms and diverse user demographics, can result in more comprehensive and valuable data suitable for annotation. Generative adversarial networks (GANs) have the potential to produce synthetic data to address dataset class imbalance challenges. In particular, novel approaches such as GAN-based oversampling (GBO) and Support Vector Machine-SMOTE-GAN (SSG) have been suggested to improve the balance of the dataset and boost the performance of the models. These artificial intelligence techniques utilize advanced neural network models such as Generative Adversarial Networks (GANs), Variational Autoencoders, and Recurrent Neural Networks to create synthetic data, addressing the difficulties associated with datasets that lack balance.

(2) *Automating Dataset Annotation.* From the findings on cyberbullying detection approaches, it is evident that these approaches rely on public datasets that are annotated. However, as the field develops, addressing the data annotation problem becomes crucial. Consequently, data annotation, such as categorizing, tagging, or labeling large volumes of raw data, poses significant challenges in constructing discriminative supervised tasks. Employing a proficient technique for automatic data annotation, as opposed to manual annotation or assigning annotators, is particularly advantageous, especially for handling large datasets. This approach not only enhances supervised learning but also reduces human effort. Therefore, investigating methods for data collection and annotation, or exploring unsupervised learning-based solutions, emerges as gaps in the area of cyberbullying detection.

(3) *Preparing Data to Ensure its Quality.* The effectiveness of cyberbullying detection methods significantly depends on the quality and availability of the data used for training, thereby influencing the resulting approaches tailored to a specific problem area. Consequently, if the data are of poor quality, showing behaviors like inconsistency, low quality, sparsity, noise, non-representativeness, ambiguous values, imbalance, and irrelevant features, the resulting cyberbullying detection approaches may exhibit reduced accuracy. Data challenges can delay effective processing and lead to inaccurate conclusions, posing a significant challenge in obtaining insights from the data. Thus, cyberbullying detection approaches must also evolve to address these emerging data challenges. Consequently, there is a need for the development of effective data pre-processing techniques tailored to the specific characteristics and challenges of the data, representing another gap in this field.

*6.3.2 Diverse Languages.* These are future studies related to language diversity issues in cyberbullying detection on social media platforms, as follows:

(1) *Advanced Natural Language Processing (NLP) Techniques.* By employing advanced Natural Language Processing (NLP) techniques such as emotion detection, sarcasm recognition, and sentiment analysis, the performance of cyberbullying detection methods can be increased. These methods empower approaches to grasp the nuances of language more adeptly, thereby improving their ability to classify cyberbullying behaviors.

(2) *Cross-Lingual Approaches.* Cross-lingual approaches enable detection algorithms to analyze instances across multiple languages, expanding the dataset used for model training. This varied dataset captures a wider range of linguistic patterns and cyberbullying tendencies, resulting in more precise detection. In addition, models trained on data from one language can be adapted to identify cyberbullying in other languages through transfer learning. This process allows algorithms to use insights from one language to improve performance in others, thus improving accuracy in various linguistic contexts. In addition, despite language differences, certain characteristics that reveal cyberbullying may be common across languages. Cross-lingual techniques identify and utilize these universal behaviors, strengthening the effectiveness of detection algorithms. Furthermore, cross-lingual methods facilitate data augmentation by utilizing labeled data from one language to refine model training in languages with limited labeled data. This augmentation enriches the training dataset, bolstering the generalization capability of detection models and heightening accuracy.

(3) *Multilingual Pre-Trained Language Models.* Multilingual pre-trained language models provide a solution for applying models to low-resource languages without requiring additional training, thereby addressing the data availability gap between high and low-resource languages. However, the training and fine-tuning processes suffer from significant computational costs as a result of the models' large parameter sizes. Enhancing the efficiency and interpretability of multilingual pre-trained language models motivates further investigation into cyberbullying instances on social media platforms. In addition, two emerging strategies aim to improve the generalizability and scalability of these models. The first approach involves pre-training models using data from relevant sources, while the second focuses on models tailored to specific low-resource languages.

(4) *Code-Switching Modeling.*The development of code-switching models will enhance the accuracy and performance of models in cyberbullying detection on social media platforms due to the ability of these models to classify and interpret cyberbullying instances in mixed-language data and handle code-switching scenarios. Sometimes, data may be found that contain more than one language in one instances, so code-switching modeling is needed to address this mixed-language challenge.

*6.3.3 Detection Approaches.* These are future studies related to the approach issues in cyberbullying detection on social media platforms as follows:

(1) Multimodal Approaches . A promising direction in cyberbullying detection lies in the adoption of multimodal approaches that integrate text with other modalities such as images, audio, and metadata. Social media content is rarely limited to textual interactions; instead, it often combines memes, videos, emojis, or user behavioral patterns that convey implicit or explicit abusive signals. Relying solely on text may, therefore, overlook critical contextual information, particularly in cases where visual or auditory elements reinforce or alter the meaning of a message. By combining these modalities, multimodal models can capture richer semantic and contextual cues, leading to improved performance and robustness.

(2) *Ensemble Methods.* Combining multiple detection techniques into an ensemble model can leverage the strengths of each approach while mitigating individual weaknesses, thus improving overall detection accuracy. Ensemble models improve predictive accuracy by aggregating predictions from several models, resulting in an improved classification of cyberbullying categories. Furthermore, ensemble models have been shown to outperform individual base models in evaluation metrics such as accuracy and precision.

(3) *Large language models (LLMs).* Introduction of large language models (LLMs) increases the efficiency in cyberbullying detection on social media platforms, including GPT-3, GPT-3.5, GPT-4, FLAN, and LIaMI. These models become more effective even with a small volume of data for training models since they deliver outstanding performance in cyberbullying detection. The state-of-the-art results of natural language processing tasks such as cyberbullying detection have been achieved by Large Language Models (LLMs), including BERT and RoBERTa. Their capacity to grasp the intricacies and context of language aids in precisely recognizing bullying instances across online communication forums and social media platforms. Furthermore, LLMs are adept at managing extensive datasets and continuously improving their performance through learning, thereby enhancing their efficacy over time.

*6.3.4 Impacts.* These are impactful approaches in cyberbullying detection on social media platforms as follows:

(1) *Improving of User Safety.* Improving cyberbullying detection methods could enhance user safety by promptly recognizing and addressing harmful conduct. This, in turn, could foster a more favorable online environment, especially for vulnerable demographics like youths and adolescents.
(2) *Considering of Ethical Issues.* The identification of cyberbullying instances relies on analyzing user data alongside personal characteristics such as name, gender, age, interests, and other human behaviors. As detection algorithms advance, social media platforms must grapple with ethical issues related to privacy, security, bias, and censorship. Striking a delicate balance between safety, user privacy, and the right to freedom of expression poses a complex challenge that requires careful deliberation.
(3) *Moderating of User-generated Content Automatically.* Automatic instances moderation encompasses the monitoring, filtering, and management of user-generated instances on social media platforms. Cyberbullying detection methods analyze user-generated instances to identify and remove material that violates community standards or legal requirements, such as offensive or abusive instances. By quickly identifying and removing inappropriate or harmful instances without human intervention, automatic instances moderation helps social media platforms maintain a safe and positive online environment.

*6.3.5 Defense.* These are technical solutions that are designed to defend against cyberbullying instances on social media platforms as follows.

(1) *Content Moderation Frameworks.* The development of advanced algorithms, which may include machine learning, deep learning, large language models, or generative approaches, will help to classify harmful flags automatically and in real-time. These instances moderation frameworks might include text-based algorithms, image and video recognition, pattern recognition algorithms, and user analysis algorithms. Advanced algorithms analyze instances in all modalities such as text, images, and videos to classify patterns that reveal cyberbullying behavior. By implementing flagging mechanisms, social media platforms may take proactive measures to combat cyberbullying instances, including selectively filtering and removing harmful instances, blocking offensive words such as insults, sarcasm, and threats, and issuing warnings to users about their behavior in platform cyberspace.
(2) *Keyword Filtering* The implementation of filtering techniques may help classify and remove abusive words or indirect forms such as sarcasm and masked insults frequently used in cyberbullying on social media platforms. These techniques might integrate keyword and natural language processing to facilitate context understanding of cyberbullying instances. Keyword filtering might include profane language and sensitive topic filters, while natural language processing might involve sentiment analysis, context analysis, named entity recognition, and topic modeling. Natural language processing algorithms may be used by social media platforms to analyze online instances posted by users, developing dedicated databases that store abusive, nuanced and offensive words, expressions, and patterns for future use as a significant way to combat cyberbullying.

*6.3.6 Mitigation.* These are protective mechanisms against cyberbullying instances on social media platforms as follows:

(1) *Additional Features.* Apart from developing various algorithms to detect and identify cyberbullying instances on social media platforms, there is a need, especially within social media platform companies, to integrate robust methods to combat cyberbullying. This includes implementing filtering mechanisms for inappropriate language, blocking disruptive users, and empowering users with control. In addition, social media platforms could introduce reporting mechanisms for code-of-conduct violations, such as doxing, impersonations, flaming, self-harm, denigration, outing, insults, and harassment on behalf of users. These measures aim to discourage online harassment behaviors and safeguard against cyberbullying instances on social media platforms.

(2) *Deactivation Methods.* Social media platforms may implement mechanisms related to reauthentication before users post any instances through their account. This reauthentication process may involve the use of mobile numbers or emails. Social media platforms randomly send verification codes to users through these methods, and if users do not verify their email or mobile number, it indicates that the account is fake and will be automatically deactivated. Additionally, during the registration process, users can undergo a mechanism to verify the information they enter, such as ensuring that names and birthdays match the information provided through their mobile number and email. If the information does not match, the system will not allow users to register, thus preventing the creation of fake accounts. This method helps protect users of social media platforms against cyberbullying instances and enhances overall safety in the cyberspace of social media platforms.

(3) *Voting Mechanism.* In social media platforms, users post different instances and may be viewed by various users if they have followed them or are friends, so they can view each other's online activities and vice versa. Through these online activities, social media platforms can integrate mechanisms for users to vote on harassment posts or comments, and when a large number of users agree that instances are harmful, they will automatically be deleted from the platforms. This mechanism will help mitigate the creation of cyberbullying instances on social media platforms.

*6.3./ Recommendations.* These are potential recommendations to combat cyberbullying instances on social media platforms as follows:

(1) *Benchmark of Data Annotation.* Using automated methods such as Self-Supervised Learning (SSL) models to label data effectively is key. Also, employing diverse data augmentation techniques like those based on GANs and autoencoders can produce annotated data for training models, thereby enhancing the accuracy of cyberbullying detection. Moreover, manual annotation by linguists and experts is essential for identifying cyberbullying factors and training models to recognize these aspects in comments, thereby improving comprehension of cyberbullying patterns. By integrating automated and manual annotation methods, the precision and effectiveness of cyberbullying detection models can be notably boosted.

(2) *Low Resource Languages.* The surge in online communication platform users across Asia and African nations has led to a corresponding rise in cyberbullying instances on social media platforms, particularly in languages such as Swahili, Hindi, Roman Urdu, Malay, Bengali, and Marathi, which lack robust linguistic support. Detecting cyberbullying in these low-resource languages poses significant challenges due to linguistic intricacies, informal expressions, and the prevalence of slang. Addressing this issue requires the development of specialized models tailored to these linguistic nuances. Such models would enable the identification of cyberbullying instances that might otherwise evade detection. Moreover, there is a pressing need to create datasets specifically for these less powerful languages, as existing public datasets are limited in their coverage.

(3) *Multi-labeled Based Classification.* Accurate detection of cyberbullying necessitates comprehension of various factors, including but not limited to gender, religious affiliation, ethnicity, age, levels of aggression, and instances of non-cyberbullying. The complexity of cyberbullying has raised concerns regarding its diverse forms. Existing approaches to cyberbullying instances classification predominantly rely on single-label datasets, categorizing instances as either cyberbullying or non-cyberbullying. However, cyberbullying encompasses a range of behaviors, such as verbal abuse, threats, and the dissemination of inappropriate material. Moreover, cyberbullying can take various linguistic forms, spanning text, images, and multimedia instances. Given this complexity, it is imperative to consider multi-labeled datasets to gauge the severity of harassment in online interactions. Multi-label datasets facilitate the categorization of social media platform posts into multiple types of bullying, allowing for a thorough examination of cyberbullying occurrences. This approach facilitates the detection of cyberbullying across different communication channels and mediums.

(4) *Deepfake Technology.* Due to the development of technology, especially the enhancement of artificial intelligence, there are now events occurring in which someone creates instances and convinces other people that the instances was created by someone else, but it is not true. This instances is often used to harass or bully another person. Deepfake instances may be categorized into audio, image, and video deepfakes. Therefore, more studies are needed in this area to combat the situation.

(5) *Interdisciplinary Research.* Due to the diversity of cyberbullying and its impacts on communities, interdisciplinary research across various fields such as computer science, sociology, psychology, political science, and management is essential. This collaborative effort will help gather more analytical data relevant to communities affected by cyberbullying, including victims, bystanders, and bullies. This broader approach will enhance our understanding of how to combat the situation by reaching people of different ages, cultures, positions, and statuses. Communities, especially those in sociology and psychology, possess significant insights into cyberbullying, as they address various mental health issues stemming from online harassment for various reasons.

## 7 CONCLUSION

This review has explored cyberbullying detection methodologies across social media platforms, focusing on machine learning, deep learning, traditional techniques, and large language models (LLMs). It has examined datasets, labeling techniques, and sources while also addressing challenges such as dataset scarcity, model generalizability, and linguistic diversity. The review has identified challenges in open source code availability and comparative studies, emphasizing the need for resources in non-English languages. Twitter has emerged as the primary dataset source due to easy accessibility, with many datasets sourced from platforms like Facebook, Instagram, Reddit, Kaggle, etc. However, public datasets often suffer from inaccuracies, imbalances, and limitations that impact detection accuracy. The SVM algorithm and TF-IDF features are widely utilized, while traditional approaches face limitations in processing large datasets and real-time detection. Deep learning methods offer promise in handling large datasets and automatic feature extraction. High-resource languages such as English are extensively studied, highlighting the importance of analyzing low-resource languages such as Swahili and Bengali. LLMs such as GPT and BERT play an important role in enhancing detection accuracy. The review highlights the need for new and unbiased datasets in all languages and improved detection methodologies. Ultimately, it emphasizes the importance of classifying cyberbullying instances across all modalities for effective detection.

## ACKNOWLEDGMENTS

This research was partly supported by the European Celtic+ and Swedish Vinnova project (C2022/1-3) "CISSAN-Collective Intelligence Supported by SecurityAware Nodes", and the Knowledge Foundation, Sweden, through the Human-Centered Intelligent Realities (HINTS) Profile Project (contract 20220068).